\pdfoutput=1

\documentclass[11pt,twoside,a4paper,cmspaper,final,collab]{cms-tdr}
\def\svnVersion{177388}\def\cmsCernNo{2012-320}\def\cmsCernDate{\today}\def\cmsMessage{Submitted to Physics Letters B}

\begin{document}\cmsNoteHeader{HIN-12-015}

\hyphenation{had-ron-i-za-tion}
\hyphenation{cal-or-i-me-ter}
\hyphenation{de-vices}

\RCS$Revision: 156919 $
\RCS$HeadURL: svn+ssh://svn.cern.ch/reps/tdr2/papers/HIN-12-015/trunk/HIN-12-015.tex $
\RCS$Id: HIN-12-015.tex 156919 2012-11-07 18:51:50Z alverson $
\newlength\cmsFigWidth
\ifthenelse{\boolean{cms@external}}{\setlength\cmsFigWidth{0.95\columnwidth}}{\setlength\cmsFigWidth{0.8\textwidth}}
\ifthenelse{\boolean{cms@external}}{\providecommand{\cmsLeft}{top}}{\providecommand{\cmsLeft}{left}}
\ifthenelse{\boolean{cms@external}}{\providecommand{\cmsRight}{bottom}}{\providecommand{\cmsRight}{right}}
\renewcommand\d{\rd}
\newcommand {\rootsNN}  {\ensuremath{\sqrt{s_{_{NN}}}}}
\newcommand {\pp}{\Pp\Pp\xspace}
\newcommand {\pPb}{\ensuremath{\text{pPb}}\xspace}
\newcommand {\roots}    {\ensuremath{\sqrt{s}}}
\newcommand {\dphi}     {\ensuremath{\Delta\phi}}
\newcommand{\noff}    {\ensuremath{N_\text{trk}^\text{offline}}\xspace}
\newcommand {\deta}     {\ensuremath{\Delta\eta}}
\newcommand {\AonA}  {\ensuremath{\mathrm{AA}}\xspace}
\newcommand {\pA}    {\ensuremath{\mathrm{pA}}\xspace}
\newcommand {\PbPb}  {\ensuremath{\text{PbPb}}\xspace}
\providecommand{\mubinv} {\ensuremath{\,\mu\mathrm{b}^{-1}}\xspace}
\providecommand{\HYDJET} {\textsc{hydjet}\xspace}
\providecommand{\AMPT} {\textsc{ampt}\xspace}
\cmsNoteHeader{HIN-12-015} % This is over-written in the CMS environment: useful as preprint no. for export versions
\title{Observation of long-range, near-side angular correlations in \pPb\  collisions at the LHC}%

\date{\today}

\abstract{
Results on two-particle angular correlations for charged particles emitted in \pPb\
collisions at a nucleon-nucleon center-of-mass energy of 5.02\TeV are presented.
The analysis uses two million collisions collected with the CMS detector at the LHC.
The correlations are studied over a broad range of
pseudorapidity, $\eta$, and full azimuth, $\phi$,
as a function of charged particle multiplicity and particle transverse
momentum, $\pt$. In high-multiplicity events, a
long-range ($2<|\deta|<4$), near-side ($\dphi\approx 0$) structure emerges in the
two-particle \deta-\dphi\ correlation functions.
This is the first observation of such
correlations in proton--nucleus collisions, resembling the ridge-like correlations
seen in high-multiplicity \pp\ collisions at $\roots = 7$\TeV and in
\AonA\ collisions over a broad range of center-of-mass energies.
The correlation strength exhibits a pronounced maximum
in the  range  of $\pt =1$--1.5\GeVc and an approximately linear increase
with charged particle multiplicity for high-multiplicity events. These observations
are qualitatively similar to those in \pp\ collisions when selecting
the same observed particle multiplicity, while the overall strength of the correlations is
significantly larger in \pPb\ collisions.}

\hypersetup{%
pdfauthor={CMS Collaboration},%
pdftitle={Observation of long-range near-side angular correlations in  proton-lead collisions at the LHC}, % Force line breaks with \\
pdfsubject={CMS},%
pdfkeywords={CMS, physics, heavy ions}}

\maketitle %maketitle comes after all the front information has been supplied
\section{Introduction}

This Letter presents measurements of two-particle angular correlations
in proton-lead (\pPb) collisions at a nucleon-nucleon center-of-mass energy $\rootsNN = 5.02$\TeV,
performed with the Compact Muon Solenoid (CMS) detector at
the Large Hadron Collider (LHC).
Two-particle correlations in high-energy collisions provide
valuable information for characterizing Quantum Chromodynamics and
have been studied  previously for a broad range of collision energies
in proton-proton (\pp), proton-nucleus (\pA),  and nucleus-nucleus (\AonA) collisions.
Such measurements can elucidate the underlying mechanism of particle
production and possible collective effects resulting from the high particle
densities accessible in these collisions.

Studies of two-particle angular correlations are typically performed using
two-dimensional $\deta$-$\dphi$ correlation functions, where
$\dphi$ is the difference in azimuthal angle $\phi$ between the two particles
and $\deta$ is the difference in pseudorapidity $\eta$ =$-\ln(\tan(\theta/2))$.
The polar angle $\theta$ is defined relative to the counterclockwise beam.

Of particular interest in studies of collective effects
is the long-range (large $|\deta|$) structure
of two-particle correlation functions, which is less susceptible to known
sources of correlations such as resonance decays and fragmentation of energetic jets.
Measurements in high-energy
\AonA\ collisions have shown significant modifications of the long-range
structure compared with minimum bias \pp\ collisions~\cite{Alver:2008gk}. Novel correlation structures
extending over large $\deta$ at $|\dphi| \approx 0$ and $|\dphi| \approx 2\pi/3$
were observed in azimuthal correlations for intermediate particle transverse momenta,
$\pt \approx 1$--5\GeVc~\cite{Adler:2005ee,Adams:2005ph,Abelev:2009af,Alver:2009id,Abelev:2009jv,
Chatrchyan:2011eka,Chatrchyan:2012wg,Aamodt:2011by,ATLAS:2012at}.
In \AonA\ collisions, long-range correlations are interpreted as a
consequence of the hydrodynamic flow of the produced strongly interacting medium \cite{Ollitrault:1992bk}
and are usually characterized by the Fourier components of the azimuthal particle distributions \cite{Voloshin:1994mz}.
Of particular  importance are the second and third Fourier components, called
elliptic and triangular flow, as they most directly reflect the
 medium response to the initial collision geometry and its fluctuations \cite{Alver:2010gr},
and allow the study of fundamental transport properties of the medium using hydrodynamic
models~\cite{Alver:2010dn,Schenke:2010rr,Qiu:2011hf}.

In current \pp\ and \pA\ Monte Carlo (MC) event generators,
the dominant sources of such long-range correlations are momentum conservation and
away-side ($\dphi \approx \pi$) jet correlations.
Measurements in \pp\ collisions at 7\TeV\ have revealed the emergence of
long-range, near-side ($\dphi \approx 0$) correlations in a selection of collisions
with very high final-state particle multiplicity~\cite{Khachatryan:2010gv}.
A large variety of theoretical models have been proposed to explain the origin of these
so-called ridge-like correlations (see Ref.~\cite{Li:2012hc} for a recent review).
The proposed mechanisms range from color connections in hard
scattering processes and collective effects in the initial interaction of the protons to
hydrodynamic effects in the high-density system possibly formed in these collisions.
It is natural to search for the possible emergence of related features in
\pPb\ collisions, where a similar range of final-state multiplicities can be explored.
The first comparison of \pp\ and \pPb\ measurements as a function of charged particle
multiplicity and particle transverse momentum, presented in this Letter, should
provide valuable information for understanding
the origin of the long-range, near-side correlation signal seen in high-multiplicity \pp\ collisions.

\section{Experimental Setup}
\label{sec:exp_evt}

This analysis uses a \pPb\ data set collected during a short run (lasting for about 8 hours) in September 2012.
The beam energies were 4\TeV for protons and 1.58\TeV per nucleon for lead nuclei,
resulting in a center-of-mass energy per nucleon pair of $\rootsNN = 5.02$\TeV.
Due to the energy difference, the  nucleon-nucleon  center-of-mass  in the \pPb\ collisions is not
at rest with respect to the laboratory frame.
Since the higher energy proton beam traveled in the clockwise direction,
i.e.\ at $\theta = \pi$,  massless particles
emitted at $\eta_\text{cm} = 0$ in the nucleon-nucleon center-of-mass
frame will be detected at $\eta =  -0.465$ in the laboratory frame.

A detailed
description of the CMS experiment can be found in Ref.~\cite{JINST}. The
main detector subsystem used for this analysis is the tracker,
located in the 3.8\unit{T} field of the superconducting solenoid.
It measures charged particles within the pseudorapidity range $\abs{\eta}< 2.5$
and consists of 1440 silicon pixel and 15\,148 silicon strip detector modules.
It provides an impact parameter resolution of ${\sim}15\mum$ and a transverse
 momentum (\pt) resolution of about 1.5\% for 100\GeVc particles.
Also located inside the solenoid are
the electromagnetic calorimeter (ECAL)  and hadron calorimeter (HCAL).
The ECAL consists of more than 75\,000
lead-tungstate crystals, arranged in a quasi-projective geometry and distributed in a
barrel region ($|\eta| < 1.48$) and two endcaps that extend up to $|\eta| = 3.0$.
The HCAL barrel and endcaps are sampling calorimeters composed of brass and
scintillator plates, covering $|\eta| < 3.0$.
Iron forward calorimeters (HF) with quartz fibers, read
out by photomultipliers, extend the calorimeter coverage up to
$|\eta| = 5.0$.  The detailed
MC simulation of the CMS detector response is based on \GEANTfour\
\cite{GEANT4}.

\section{Event and Track Selection}

The relatively low \pPb\ collision frequency (about 200\unit{Hz}) provided by the LHC in this pilot run
allowed the use of a track-based minimum bias trigger.
For every \pPb\ bunch crossing, the detector
was read out and events were accepted if at least
one track with $\pt > 0.4$\GeVc was found in the pixel tracker.
In the offline analysis, a coincidence of at least one HF calorimeter tower
with more than 3\GeV of total energy on both the positive and negative sides of HF
was required to select hadronic collisions.
Events were also required to contain at least one reconstructed primary
vertex within 15\unit{cm} of the nominal interaction point along the beam axis ($z_\text{vtx}$)
and within 0.15\unit{cm} transverse distance to the beam trajectory.
At least two reconstructed tracks were required to be associated
with the primary vertex. Beam related background was suppressed by rejecting events with
a high fraction of pixel clusters inconsistent with a single collision
vertex~\cite{Collaboration:2010xs}. Based on simulations using the \HIJING\ event generator,
the event selections have a total acceptance of about 96.2\% for hadronic inelastic pPb interactions.
A total of 2 million events passed all selection criteria,
corresponding to an integrated luminosity of about $1\mubinv$, assuming a \pPb\
interaction cross section of 2.1~barns.

The angular correlation functions were obtained using the
CMS \textit{highPurity}~\cite{CMSTracking}
track selection. Additionally, a reconstructed track was only considered as a
primary-track candidate if the significance of the separation along the beam axis,
$z$, between the track and the primary vertex, $dz/\sigma(dz)$, and the significance
of the impact parameter relative to the primary vertex transverse to the beam,
$dT/\sigma(dT)$, were each less than 3. The relative uncertainty
of the momentum measurement, $\sigma(\pt)/\pt$, was required to be less than 10\%.
To ensure high tracking efficiency and low fake rate, only tracks within $|\eta|<2.4$
and with $\pt >$ 0.1\GeVc were used.

To match the analysis used for high-multiplicity
\pp\ collisions \cite{Khachatryan:2010gv}, the events were divided into classes
of reconstructed track multiplicity, \noff,
where primary tracks with $|\eta|<2.4$ and $\pt >0.4$\GeVc were counted.
The fraction of events falling into each of the four multiplicity classes
is listed in Table~\ref{tab:multbinning}.
The table also lists the average values of \noff\
and $N_\mathrm{trk}^\mathrm{corrected}$, the event
multiplicity of charged particles with $|\eta|<2.4$ and $\pt >0.4$\GeVc corrected
for detector acceptance and efficiency of the track reconstruction algorithm,
as discussed in the following section.

\begin{table}[htb]%\renewcommand{\arraystretch}{1.2}\addtolength{\tabcolsep}{-1pt}
\centering
\topcaption{\label{tab:multbinning} Fraction of the full event sample
for each multiplicity class. The last two columns show the observed
and corrected multiplicities, respectively, of charged particles with $|\eta|<2.4$ and $\pt >0.4$\GeVc.
Systematic uncertainties are given for the corrected multiplicities, while statistical uncertainties are negligible.
}
\small
\begin{tabular}{ l  c  c  c}
\hline
Multiplicity class & Fraction  & $\left<\noff \right>$ & $\left<N_\text{trk}^\text{corrected}\right>$\\
 \multicolumn{1}{c}{(\noff)}&(\%)&&\\
\hline
Minimum Bias        &   100.0  &  40.6   & 53.4$\pm$2.9\\
$ \noff <35$        &   50.4   &  17.1   & 23.5$\pm$1.3\\
$35 \leq \noff <90$ &   41.9   &  56.3   & 75.6$\pm$4.1\\
$90 \leq \noff <110$&   4.6    &  98.2   & 114.3$\pm$6.2\\
$\noff \geq 110$    &   3.1    &  128.2  & 149.1$\pm$8.1\\
\hline
\end{tabular}
\end{table}

\section{Calculation of the Two-Particle Correlation Function}
\label{sec:inclusive_analysis}

The analysis of two-particle
correlations was performed in classes of track multiplicity, \noff,
following the procedure established in
\cite{Chatrchyan:2011eka,Chatrchyan:2012wg}.
For each track
multiplicity class, ``trigger'' particles are defined as charged particles originating
from the primary vertex within a given \pt\ range.
The number of trigger particles in the event is denoted by $N_\text{trig}$.
In this analysis, particle pairs are formed by associating every trigger particle with
the remaining charged primary particles from the same \pt\ interval
as the trigger particle (a minimum of two particles is required in each \pt\ bin
from each event). The per-trigger-particle associated yield is defined as
\begin{equation}
\label{2pcorr_incl}
\frac{1}{N_\text{trig}}\frac{\d^{2}N^\text{pair}}{\d\Delta\eta\, \d\Delta\phi}
= B(0,0)\times\frac{S(\Delta\eta,\Delta\phi)}{B(\Delta\eta,\Delta\phi)},
\end{equation}
where $\Delta\eta$ and $\Delta\phi$ are the differences in $\eta$
and $\phi$ of the pair. The signal distribution, $S(\Delta\eta,\Delta\phi)$, is
the per-trigger-particle yield of particle pairs from the same event,
\begin{equation}
\label{eq:signal}
S(\Delta\eta,\Delta\phi) = \frac{1}{N_\text{trig}}\frac{\d^{2}N^\text{same}}{\d\Delta\eta\, \d\Delta\phi}\ .
\end{equation}
The mixed-event background distribution, used to account
for random combinatorial background and pair-acceptance effects,
\begin{equation}
\label{eq:background}
B(\Delta\eta,\Delta\phi) = \frac{1}{N_\text{trig}}\frac{\d^{2}N^\text{mix}}{\d\Delta\eta\, \d\Delta\phi}\ ,
\end{equation}
is constructed by pairing the trigger particles in each event with the
associated particles from 10 different random events in the same 2\unit{cm} wide $z_\text{vtx}$ range.
The symbol $N^\text{mix}$ denotes the number of pairs taken from the mixed event,
while $B(0,0)$ represents the mixed-event associated yield for
both particles of the pair going in approximately the same direction and
thus having full pair acceptance (with a bin width of 0.3 in $\Delta\eta$ and $\pi/16$ in $\Delta\phi$).
Therefore, the ratio $B(0,0)/B(\Delta\eta,\Delta\phi)$
is the pair-acceptance correction factor used to derive the corrected
per-trigger-particle associated yield distribution. The signal and background distributions
are first calculated for each event, and then averaged over all the events within
the track multiplicity class.

Each reconstructed track is weighted by the inverse of an
efficiency factor, which accounts for the detector
acceptance, the reconstruction efficiency,
and the fraction of misreconstructed tracks.
Detailed studies of tracking efficiencies using
MC simulations and data-based methods can be found in \cite{TRK-10-002}.
The combined geometrical acceptance
and efficiency for track reconstruction exceeds 50\%
for $\pt \approx 0.1$\GeVc and $|\eta|<2.4$.
The efficiency is greater than 90\% in the $|\eta|<1$ region
for $\pt>0.6\GeVc$.
For the multiplicity range studied here, little or no dependence of the tracking efficiency on
multiplicity is found and the rate of misreconstructed tracks
remains at the 1--2\% level.

Simulations of \pp, \pPb\ and peripheral \PbPb\ collisions using the \PYTHIA, \HIJING and \HYDJET event
generators, respectively, yield efficiency correction factors that vary due to the
different kinematic and mass distributions for the particles produced in these generators. Applying
the resulting correction factors from one of the generators to simulated data from one
of the others gives associated yield distributions that agree within 5\%.
Systematic uncertainties due to track quality cuts and potential contributions from secondary
particles (including those from weak decays) are examined by loosening or tightening the
track selections on $dz/\sigma(dz)$ and $dT/\sigma(dT)$ from 2 to 5.
The associated yields are found to be insensitive to these track selections within 2\%.

\begin{figure}[htbp]
\includegraphics[width=0.45\textwidth]{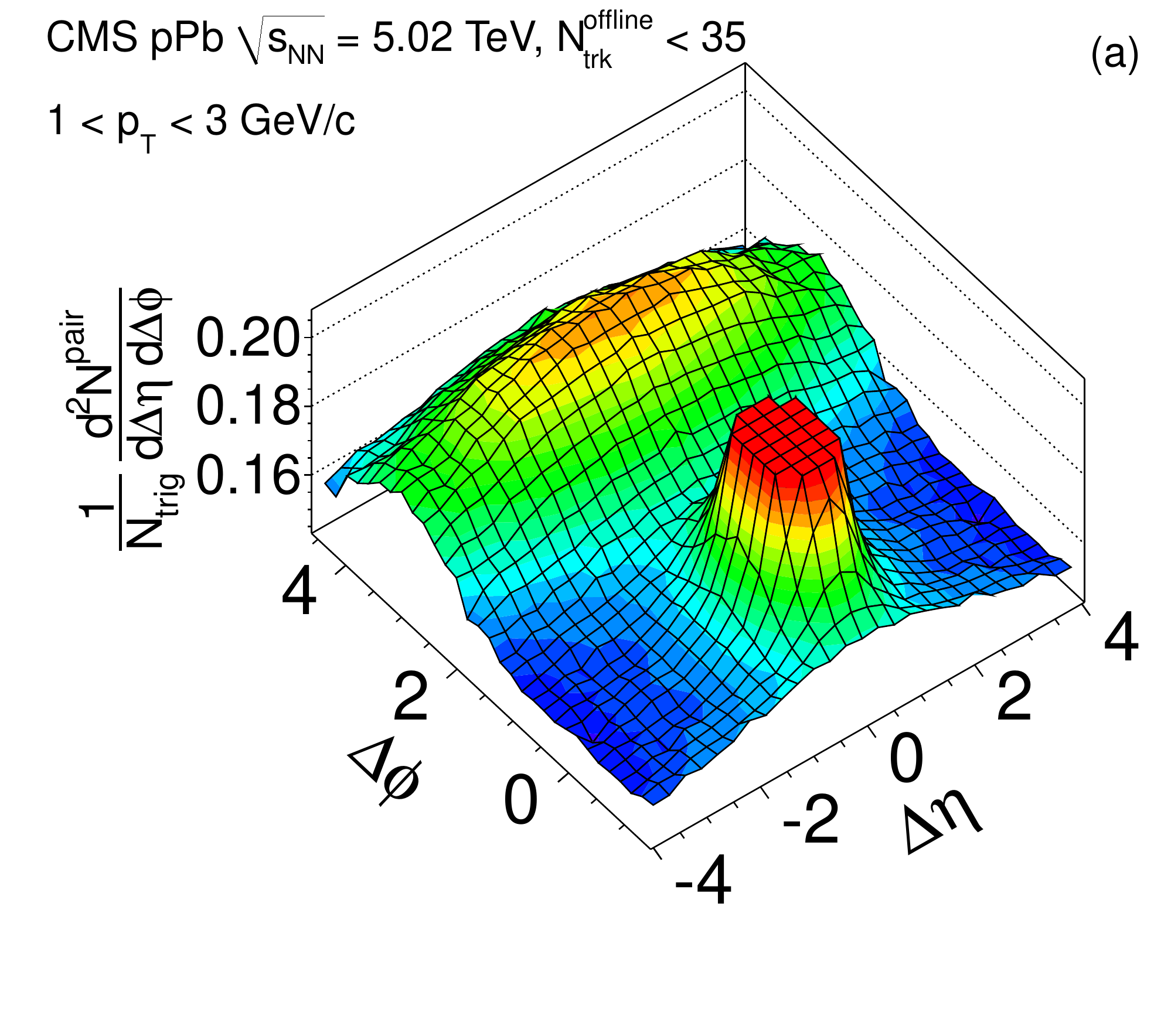}
\includegraphics[width=0.45\textwidth]{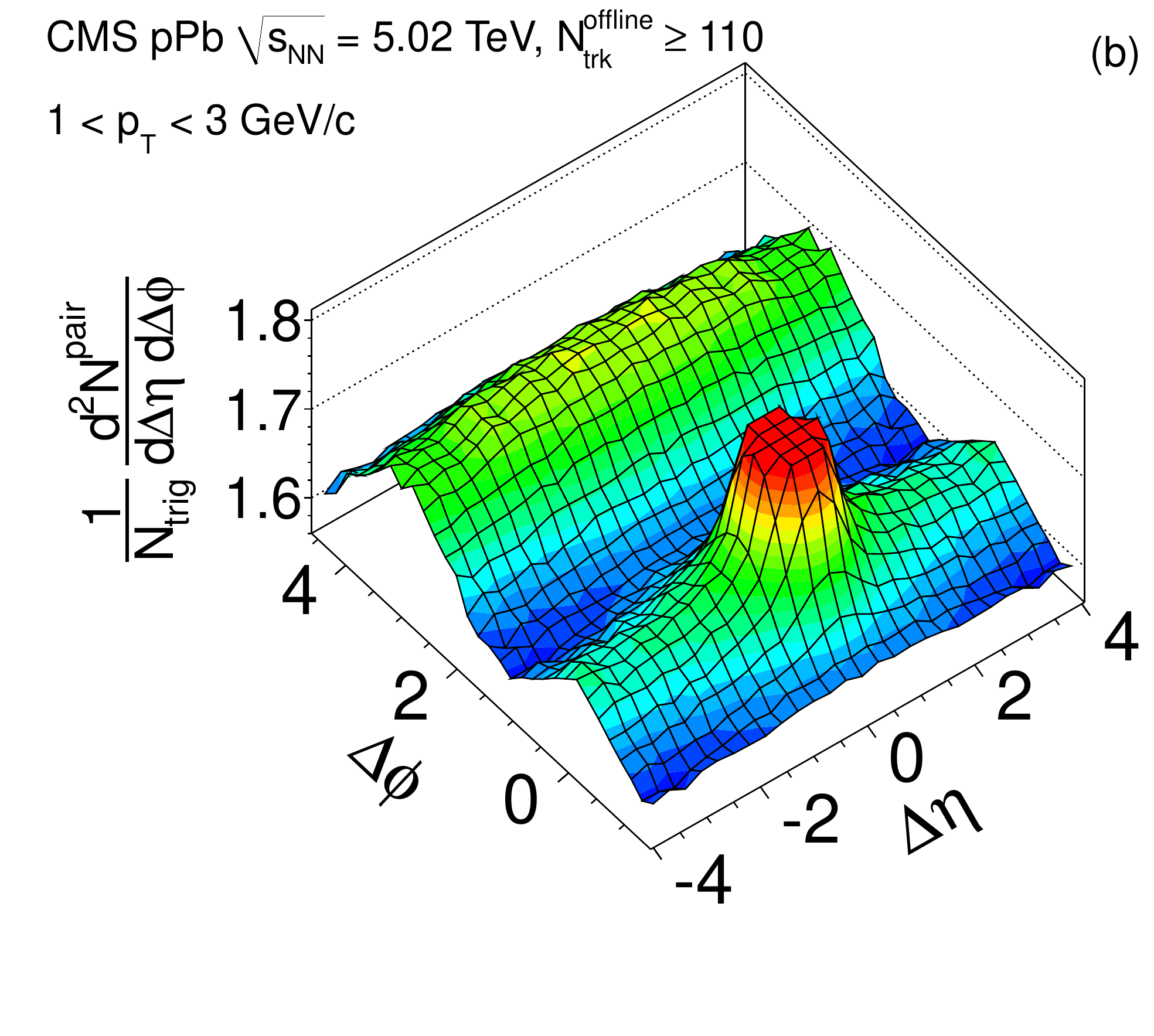}
  \caption{ \label{fig:corr2D_pPb_pt1-3} 2-D two-particle correlation
functions for 5.02\TeV \pPb\ collisions for pairs of charged particles with $1<\pt<3$\GeVc.
Results are shown (a) for low-multiplicity events ($\noff < 35$)  and (b) for a
high-multiplicity selection ($\noff \geq 110$).
The sharp near-side peaks from jet correlations have been truncated to better illustrate the
structure outside that region.
   }
\end{figure}

\section{Results}
\label{sec:highmult_results}
Figure~\ref{fig:corr2D_pPb_pt1-3} compares 2-D two-particle correlation
functions for events with low (a) and high (b) multiplicity,
for pairs of charged particles with $1<\pt<3$\GeVc. For the low-multiplicity selection ($\noff < 35$), the
dominant features are the correlation peak near $(\deta, \dphi) = (0, 0)$ for pairs of particles originating
from the same jet and the elongated structure at $\dphi \approx \pi$ for pairs of particles
from back-to-back jets. To better illustrate the full correlation structure, the jet peak has been truncated.
High-multiplicity events ($\noff \geq 110$) also show the same-side jet peak
and back-to-back correlation structures. However,
in addition, a pronounced ``ridge''-like structure emerges at $\dphi \approx 0$
extending to $|\deta|$ of at least 4 units. This observed structure is similar to that seen in
high-multiplicity \pp\ collision data at $\roots = 7$\TeV \cite{Khachatryan:2010gv} and in \AonA collisions
over a wide range of energies~\cite{Adams:2005ph,Abelev:2009af,Alver:2009id,Abelev:2009jv,
Chatrchyan:2011eka,Chatrchyan:2012wg,Aamodt:2011by,ATLAS:2012at}.

\begin{figure*}[thbp]
\centering
\includegraphics[width=0.95\linewidth]{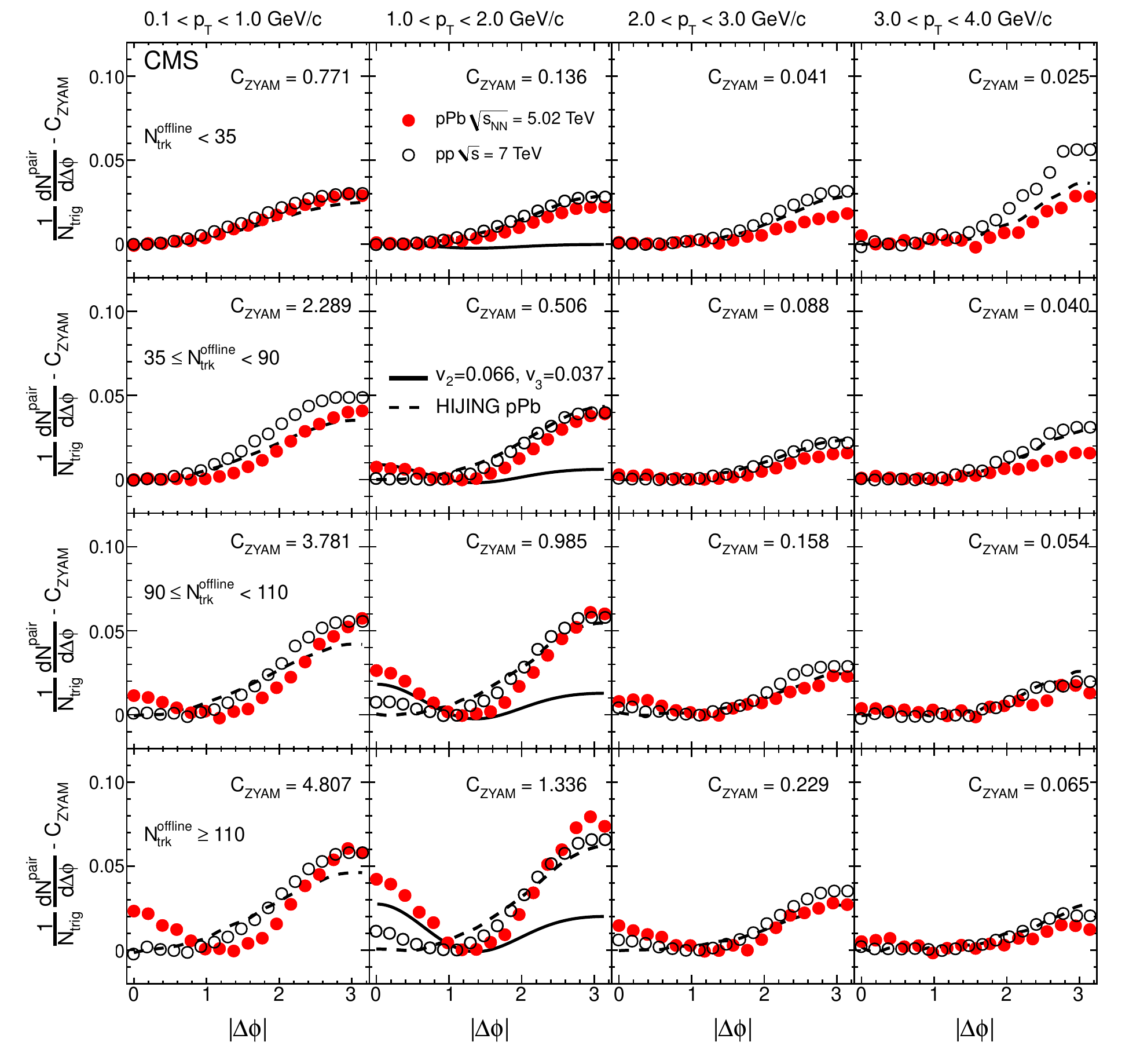}

  \caption{ \label{fig:compare_corr_phi_PAData_Minbias_5TeV_dphi2-4_highmult_4by4}
Correlated yield obtained from the ZYAM procedure as a function of $|\dphi|$ averaged over $2<|\deta|<4$ in different
\pt\ and multiplicity bins for 5.02\TeV \pPb\ data (solid circles) and 7\TeV \pp\ data (open circles).
The \pt\ selection applies to both particles in each pair. Statistical uncertainties are smaller than the marker size.
The subtracted ZYAM constant is listed in each panel.
Also shown are \pPb\ predictions for \HIJING~\cite{Gyulassy:1994ew} (dashed curves) and a
hydrodynamic model~\cite{Bozek:2011if} (solid curves shown for $ 1 < \pt < 2$\GeVc).
   }
\end{figure*}

As a cross-check, correlation functions were also generated for tracks paired with ECAL
photons, which originate primarily from decays of $\pi^0$s, and for pairs of ECAL photons. These
distributions showed similar features as those seen  in Fig.~\ref{fig:corr2D_pPb_pt1-3}, in particular the
ridge-like correlation for high multiplicity events.

To investigate the long-range, near-side correlations
in finer detail, and to provide a quantitative comparison to \pp\ results,
one-dimensional (1-D) distributions in $\Delta\phi$ are found
by averaging the signal and background two-dimensional (2-D) distributions over
$2 < |\deta| < 4$~\cite{Khachatryan:2010gv,Chatrchyan:2011eka,Chatrchyan:2012wg}.
In the presence of multiple sources of correlations,
the yield for the correlation of interest is commonly estimated using an implementation of the
zero-yield-at-minimum (ZYAM) method~\cite{PhysRevC.72.011902}.
A second-order polynomial is first fitted to the 1-D $\Delta\phi$ correlation function
in the region $0.1<|\Delta\phi|<2$.
The minimum value of the polynomial, $C_\mathrm{ZYAM}$, is then subtracted from the
1-D \dphi\ correlation function
as a constant background (containing no information about correlations) to shift its
minimum to be at zero associated yield.
The statistical uncertainty on the minimum level of
$\frac{1}{N_\text{trig}}\frac{\d N^\text{pair}}{\d\Delta\phi}$
obtained by the ZYAM procedure as well as the deviations found
by varying the fit range in $\Delta\phi$ give an
absolute uncertainty of ${\pm}0.0015$ on the associated yield, independent of multiplicity
and \pt.

Figure~\ref{fig:compare_corr_phi_PAData_Minbias_5TeV_dphi2-4_highmult_4by4}
shows the results for
\pPb\ data (solid circles) for various selections in \pt\ and multiplicity \noff, with \pt\ increasing from left to right
and multiplicity increasing from top to bottom.
The results for \pp\ data at $\roots = 7$\TeV,
obtained using the same procedure \cite{Khachatryan:2010gv}, are also plotted (open circles).

A clear evolution of the $\dphi$ correlation
function as a function of both \pt\ and \noff\ is observed. For the lowest multiplicity selection in \pp\ and \pPb\ the correlation
functions have a minimum at $\dphi = 0$ and a maximum at $\dphi = \pi$, reflecting the correlations from
momentum conservation and the increasing  contribution from back-to-back jet-like correlations at higher \pt.
Results from the \HIJING~\cite{Gyulassy:1994ew} model (version 1.383), shown as dashed lines, qualitatively
reproduce the shape of the correlation function for low \noff.

For multiplicities $\noff \geq 35$, a second local
maximum near $|\dphi| \approx 0$ emerges in the \pPb\ data, corresponding to the near-side, long-range ridge-like structure. In \pp\  data,
this second maximum is clearly visible only for $\noff > 90$.
For both \pp\ and \pPb\ collisions, this near-side correlated yield is largest in the $1 < \pt < 2$\GeVc range
and increases with increasing multiplicity.
While the
evolution of the correlation function is qualitatively similar in \pp\ and \pPb\ data, the absolute
near-side correlated yield is significantly larger in the \pPb\ case.

In contrast to the data, the \HIJING calculations show a correlated yield of zero at $\dphi  = 0$ for
all multiplicity and \pt\ selections. The  long-range, near-side enhancement is also absent in
simulated \pp\ collision events
with the \PYTHIA~\cite{Sjostrand:2006za,Sjostrand:2007gs} event generator (version 6.4.24)
and in simulated \pPb\ collisions with the \AMPT~\cite{Lin:2004en} model (version 1.25/2.25).

Long-range correlations in \pPb\ collisions have been quantitatively predicted in models assuming a
collective hydrodynamic expansion of a system with fluctuating initial conditions~\cite{Bozek:2011if}.
The correlation resulting from the predicted elliptic and triangular flow
components for \pPb\ collisions  at \rootsNN\ = 4.4\TeV are compared to the observed correlation
in Fig.~\ref{fig:compare_corr_phi_PAData_Minbias_5TeV_dphi2-4_highmult_4by4} for the
$1<\pt <2$\GeVc\ selection (solid line, second column).
The magnitudes for elliptic and triangular flow of $v_2 = 0.066$ and $v_3 = 0.037$
correspond to those given in Ref.~\cite{Bozek:2011if} for the highest multiplicity selection
and the average value of \pt\ $\approx$ 1.4\GeVc found in the data. The same $v_2$ and $v_3$
coefficients were used for all multiplicity  classes, showing the multiplicity dependence
of the correlated yield assuming a constant flow effect.
While this provides an indicative and useful illustration of the magnitude of the observed near-side enhancement,
a detailed quantitative comparison of the model and data will need to include the additional
non-hydrodynamical correlations from back-to-back jets, as well as the effects of momentum conservation,
which suppress the correlation near $\Delta\phi \approx 0$ relative to $\Delta\phi \approx \pi$.

The ridge-like structure in \pPb\ collisions was also predicted to arise from initial
state gluon correlations in the color-glass condensate framework, where the contribution
of collimated gluon emissions is significantly enhanced in the gluon
saturation regime~\cite{Dusling:2012cg}. This model
qualitatively predicts the increase in the correlation strength for higher multiplicity \pPb\
collisions, although it remains to be seen if the large associated yield seen in the highest
multiplicity selection can be quantitatively reproduced in the calculation.

\begin{figure*}[thb]
\centering
\includegraphics[width=0.95\linewidth]{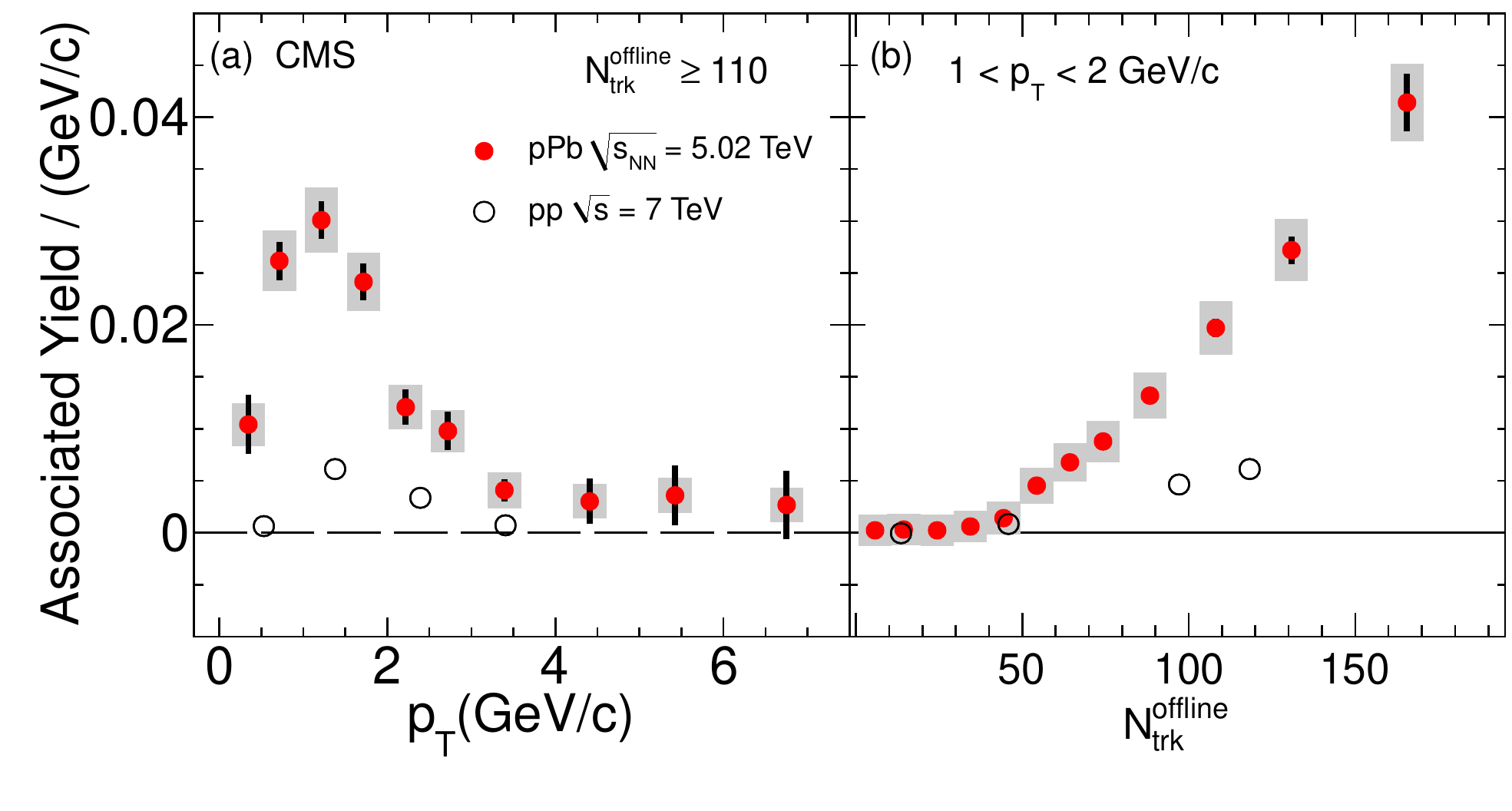}
  \caption{ \label{fig:yieldvsptvsmult_paper} Associated yield for the near-side
  of the correlation function averaged over $2<|\deta|<4$ and integrated over the region $|\dphi| < 1.2$
  in 7\TeV \pp\ collisions (open circles) and 5.02\TeV \pPb\ collisions
(solid circles). Panel (a) shows the associated yield as a function of
\pt\ for events with $\noff \geq 110$.
In panel (b) the associated yield for $1 < \pt < 2$\GeVc\ is shown as a function
of multiplicity \noff. The \pt\ selection applies to both particles
in each pair.
The error bars correspond to statistical uncertainties,
while the shaded areas denote the systematic uncertainties.
   }
\end{figure*}

The strength of the long-range, near-side correlations
can be further quantified by integrating the correlated yield from Fig.~\ref{fig:compare_corr_phi_PAData_Minbias_5TeV_dphi2-4_highmult_4by4}
over $|\dphi| < 1.2$ using 12 classes of multiplicity. The resulting integrated ``ridge yield", normalized by
the width of the \pt\ interval, is
plotted as a function of particle \pt\ and event multiplicity in Fig.~\ref{fig:yieldvsptvsmult_paper}
for \pp\ (open circles) and \pPb\ (solid circles) data.
The error bars correspond to statistical uncertainties, while the shaded boxes indicate
the systematic uncertainties.

Figure~\ref{fig:yieldvsptvsmult_paper}(a) shows that the ridge yield for events with $\noff \geq 110$
peaks in the region  $1 < \pt < 2$\GeVc for both collision systems.
However, while the yield in \pp\ collisions is consistent with zero for
the $ 0.1 < \pt < 1$\GeVc selection, it remains greater than zero in \pPb\ data
even for the $0.1 < \pt < 0.5$\GeVc\ range.
For higher \pt, the ridge yield in \pp\ collisions is consistent with zero for $\pt > 3$\GeVc ,
while the \pPb\ results only approach zero at $\pt \approx 4$--7\GeVc.

The multiplicity dependence of the ridge yield for $1 < \pt < 2$\GeVc particle pairs
is shown in Fig.~\ref{fig:yieldvsptvsmult_paper}(b).
For low-multiplicity collisions, the ridge yield determined by the ZYAM procedure
is consistent with zero, indicating that ridge-like correlations are absent or smaller than
the negative correlations expected due to, e.g.\ momentum conservation. At higher multiplicity the
ridge-like correlations emerge, with an approximately linear
rise of the ridge yield  observed for
 $\noff \gtrsim 40$, which corresponds to $N_\mathrm{trk}^\mathrm{corrected} \gtrsim 53$.
While the multiplicity dependence is qualitatively similar for \pp\ and \pPb\
collisions, a significantly larger yield per trigger particle is seen in \pPb\ than in \pp\ at a given multiplicity.

When interpreting the differences in the correlation structure between the two collision systems, it is
important to consider the relative contributions of different particle production mechanisms to
the observed particle yields. While
very high multiplicity \pp\ collisions should mainly arise from rare multiple hard-scattering processes,
the high-multiplicity \pPb\ events should mostly result from particle production in multiple soft proton-nucleon
scatterings. This will in particular affect the correlations due to back-to-back jet fragmentation in the
$\dphi \approx \pi$ region. A simultaneous description of the measurements in \pp\ and \pPb\
should provide significant constraints on models of the underlying physics processes.
With such an improved understanding of the smaller systems, comparisons to the \PbPb\
data will also provide insights in understanding the similarity of the ridge-like correlations
in all three systems.

\section{Conclusion}
\label{sec:conclusion}
The CMS detector at the LHC has been used to measure angular correlations
between two charged particles with $|\eta| <2.4$ in \pPb\ collisions at $\rootsNN = 5.02$\TeV.
Azimuthal correlations for $2 < |\deta| < 4$ in high-multiplicity \pPb\ collisions
exhibit a long-range structure at the near side ($\dphi \approx 0$).
This ridge-like structure is qualitatively similar to that observed in \pp\ collisions at \roots\ = 7\TeV
and in \AonA\ collisions over a broad range of center-of-mass energies.
The effect is most evident in the intermediate transverse momentum range, $1 < \pt < 1.5$\GeVc.
The near-side ridge yield obtained by the ZYAM procedure is found to be consistent
with zero in the low-multiplicity region, with an approximately linear increase
with multiplicity for $\noff \gtrsim 40$
(corresponding to $N_\mathrm{trk}^\mathrm{corrected} \gtrsim 53$).
While the multiplicity and \pt\ dependences of the observed effect are
similar to those seen in \pp\ data  at \roots\ = 7\TeV,
the absolute ridge yield in \pPb\ is
significantly larger than in \pp\ collisions of
the same particle multiplicity.

\section*{Acknowledgments}
We congratulate our colleagues in the CERN accelerator departments for the excellent performance of the LHC and thank the technical and administrative staffs at CERN and at other CMS institutes for their contributions to the success of the CMS effort. In addition, we gratefully acknowledge the computing centres and personnel of the Worldwide LHC Computing Grid for delivering so effectively the computing infrastructure essential to our analyses. Finally, we acknowledge the enduring support for the construction and operation of the LHC and the CMS detector provided by the following funding agencies: BMWF and FWF (Austria); FNRS and FWO (Belgium); CNPq, CAPES, FAPERJ, and FAPESP (Brazil); MEYS (Bulgaria); CERN; CAS, MoST, and NSFC (China); COLCIENCIAS (Colombia); MSES (Croatia); RPF (Cyprus); MoER, SF0690030s09 and ERDF (Estonia); Academy of Finland, MEC, and HIP (Finland); CEA and CNRS/IN2P3 (France); BMBF, DFG, and HGF (Germany); GSRT (Greece); OTKA and NKTH (Hungary); DAE and DST (India); IPM (Iran); SFI (Ireland); INFN (Italy); NRF and WCU (Korea); LAS (Lithuania); CINVESTAV, CONACYT, SEP, and UASLP-FAI (Mexico); MSI (New Zealand); PAEC (Pakistan); MSHE and NSC (Poland); FCT (Portugal); JINR (Armenia, Belarus, Georgia, Ukraine, Uzbekistan); MON, RosAtom, RAS and RFBR (Russia); MSTD (Serbia); SEIDI and CPAN (Spain); Swiss Funding Agencies (Switzerland); NSC (Taipei); ThEP, IPST and NECTEC (Thailand); TUBITAK and TAEK (Turkey); NASU (Ukraine); STFC (United Kingdom); DOE and NSF (USA).

\bibliography{auto_generated}   % will be created by the tdr script.

\cleardoublepage \appendix\section{The CMS Collaboration \label{app:collab}}\begin{sloppypar}\hyphenpenalty=5000\widowpenalty=500\clubpenalty=5000\textbf{Yerevan Physics Institute,  Yerevan,  Armenia}\\*[0pt]
S.~Chatrchyan, V.~Khachatryan, A.M.~Sirunyan, A.~Tumasyan
\vskip\cmsinstskip
\textbf{Institut f\"{u}r Hochenergiephysik der OeAW,  Wien,  Austria}\\*[0pt]
W.~Adam, E.~Aguilo, T.~Bergauer, M.~Dragicevic, J.~Er\"{o}, C.~Fabjan\cmsAuthorMark{1}, M.~Friedl, R.~Fr\"{u}hwirth\cmsAuthorMark{1}, V.M.~Ghete, N.~H\"{o}rmann, J.~Hrubec, M.~Jeitler\cmsAuthorMark{1}, W.~Kiesenhofer, V.~Kn\"{u}nz, M.~Krammer\cmsAuthorMark{1}, I.~Kr\"{a}tschmer, D.~Liko, I.~Mikulec, M.~Pernicka$^{\textrm{\dag}}$, D.~Rabady\cmsAuthorMark{2}, B.~Rahbaran, C.~Rohringer, H.~Rohringer, R.~Sch\"{o}fbeck, J.~Strauss, A.~Taurok, W.~Waltenberger, C.-E.~Wulz\cmsAuthorMark{1}
% ...
\vskip\cmsinstskip
\dag:~Deceased\\
1:~~Also at Vienna University of Technology, Vienna, Austria\\
2:~~Also at CERN, European Organization for Nuclear Research, Geneva, Switzerland\\
% ...

\end{sloppypar}
\end{document}